\documentclass[fleqn,usenatbib]{mnras}

\usepackage{newtxtext,newtxmath}
\usepackage[T1]{fontenc}

\DeclareRobustCommand{\VAN}[3]{#2}
\let\VANthebibliography\thebibliography
\def\thebibliography{\DeclareRobustCommand{\VAN}[3]{##3}\VANthebibliography}

\usepackage{subcaption}

\usepackage{graphicx}	% Including figure files
\usepackage{amsmath}	% Advanced maths commands

\title[Resizing the giants]{Resizing the giants: how modelling adiabatic interiors impacts predicted planetary radii}

\author[S. Müller \& R. Helled]{
Simon Müller,$^{1}$\thanks{E-mail: simonandres.mueller@uzh.ch}
and Ravit Helled$^{1}$
\\
$^{1}$Department of Astrophysics, University of Zürich, \\
    Winterthurerstrasse 190, 8057 Zürich, Switzerland
}

\date{Accepted 2026 April 16. Received 2026 April 6.}

\pubyear{2026}

\begin{document}
\label{firstpage}
\pagerange{\pageref{firstpage}--\pageref{lastpage}}
\maketitle

\begin{abstract}
The interiors of giant planets are commonly assumed to be convective and adiabatic, making the adiabatic temperature gradient a key ingredient in interior and evolution models. Multiple numerically distinct methods exist for computing this gradient, yet their impact on inferred planetary structure and radius has not been systematically assessed. 
In this letter we investigate how the numerical treatment of adiabatic temperature profiles affects inferred planetary radii and internal structure, comparing different methods for evaluating the adiabatic gradient against a ground-truth isentropic baseline, for both the logarithmic and non-logarithmic forms of the temperature differential equation. Static interior models of a one Jupiter mass planet were computed using a state-of-the-art hydrogen–helium equation of state.
We find that the choice of numerical method significantly impacts the inferred interior structure and radius. Using the logarithmic temperature equation, central temperatures deviate by several thousand kelvin and surface radii differ by up to 3.4 per cent, exceeding the 1 per cent precision of current giant exoplanet radius measurements threefold. The non-logarithmic form reduces deviations to below $\sim$1 per cent for most methods. We recommend spline derivatives to evaluate the adiabatic gradient, combined with the non-logarithmic temperature equation. Finite differencing and direct use of tabulated gradients or derivatives should be avoided.
\end{abstract}

\begin{keywords}
methods: numerical --
planets and satellites: gaseous planets --
planets and satellites: general --
planets and satellites: interiors 
\end{keywords}

%%%%%%%%%%%%%%%%%%%%%%%%%%%%%%%%%%%%%%%%%%%%%%%%%%

%%%%%%%%%%%%%%%%% BODY OF PAPER %%%%%%%%%%%%%%%%%%

\section{Introduction}\label{sec:introduction}

Giant planets are key objects for understanding planetary system formation and evolution, and their interior structure provides crucial constraints on their composition, thermal state, and history \citep[][]{Guillot1999,burrows_theory_2001,Fortney2010,2019ApJ...874L..31T,Muller2023,helled_giant_2026}. Accurately characterising giant planet interiors therefore requires reliable structural models. A common and well-established assumption is that giant planet interiors are largely convective, leading to adiabatic (isentropic) temperature profiles \citep[e.g.,][]{Stevenson1982,Guillot2005,Helled2014}. The adiabatic temperature gradient thus plays a central role in both static interior models and long-term evolution calculations. However, despite being mathematically equivalent in the continuous limit, different numerical implementations of the adiabatic gradient can yield meaningfully different results, and the sensitivity of interior models to these numerical choices has not been systematically explored. In this work, we investigate and quantify the differences in the planetary interiors and radii that arise when using different methods of modelling adiabatic interiors of giant planets.

This letter is structured as follows. In Section \ref{sec:methods}, we review the planetary structure equations and describe several methods of how the adiabatic temperature gradient can be calculated. We present the results of our models in Section \ref{sec:results}, where we first focus on the form of the adiabatic gradient (Section \ref{sec:adiabatic_gradient}) and then on the form of the differential equation for the temperature (Section \ref{sec:temperature_equation}). Based on these results we present practical recommendations for calculating adiabatic interior models in Section \ref{sec:recommendations}. Our main findings are summarised in Section \ref{sec:conclusions}. Finally, additional models that test the influence of the integration method and the planetary surface temperature are presented in Appendices \ref{sec:solvers} and \ref{sec:surface_temperatures}.

\section{Methods}\label{sec:methods}

Giant planets are commonly modelled as a spherically symmetric objects in hydrostatic equilibrium. Under these assumptions the structure equations are \citep[e.g..,][]{Kippenhahn2012}:

\begin{align}
        &\frac{\partial r}{\partial m} = \frac{1}{4 \pi r^2 \rho} \, , \label{eq:mass_conservation}\\ 
        &\frac{\partial p}{\partial m} = -\frac{\rm{G} m}{4 \pi r^4} \, , \label{eq:momentum_conservation}\\ 
        &\frac{\partial \log T}{\partial \log p} = \rm{min}\left(\nabla_{\rm{ad}}, \, \nabla_{\rm{rad}}\right) \, , \label{eq:diff_eq_log}
\end{align}

\noindent
where $r$, $m$, $\rho$, $p$ and $T$ are the radius, mass, density, pressure and temperature, G is the gravitational constant, and $\nabla_{\rm{rad, ad}}$ are the radiative and adiabatic temperature gradients. These equations are solved together with an appropriate equation of state and boundary conditions. Equation \ref{eq:diff_eq_log} further assumes the simple case\footnote{While more sophisticated models use the mixing length theory \citep{1958ZA.....46..108B} to determine the temperature gradient in convective regions, the adiabatic gradient still plays a key role \citep[e.g.,][]{Kippenhahn2012}.} that the temperature gradient inside the planet is the minimum of the radiative or the adiabatic gradient, which is equivalent to the Schwarzschild criterion \citep{schwarzschild_equilibrium_1906}. In this work, we focus solely on fully convective interiors, such that the temperature gradient is always the adiabatic one. The form of $\nabla_{\rm{ad}}$ commonly used in stellar and planetary structure calculations is the logarithmic temperature derivative with respect to pressure at constant entropy $S$ \citep[e.g.,][]{lamers_understanding_2017}:

\begin{equation}
    \nabla_{\rm{ad}} \equiv \left.\frac{\partial \log T}{\partial \log p} \right\vert_{S} \, ,
    \label{eq:nabla_ad}
\end{equation}

\noindent
where $S$ is the entropy. Since the adiabatic gradient is a material property, calculating it requires an appropriate equation of state. The equation of state is generally given in a tabulated form using either the thermodynamic basis $(p, T)$ or $(\rho ,T)$, and is commonly evaluated with cubic splines since derivatives can be required (for example, when using the Henyey method to solve the evolution equations \citep{Henyey1965}). For simplicity and without loss of generality, we here focus on hydrogen-helium mixtures and do not include heavy elements. 

The state-of-the-art hydrogen-helium equation of state for modelling giant planets and brown dwarfs are the tables from \citet{2021ApJ...917....4C} (hereafter CD21). These tables provide the entropy, logarithmic entropy derivatives with respect to pressure and temperature, and the adiabatic gradient as a function of the pressure-temperature or density-temperature thermodynamic bases. There are tables for pure hydrogen and helium, or hydrogen-helium mixtures for a few values of the helium mass fraction. The hydrogen-helium mixture tables offer 
(among others\footnote{Using Maxwell relations and the triple-product-rule, there are many ways deriving expressions for the adiabatic temperature gradient; see, e.g., \citet{1994sipp.book.....H} or Equation 12 in \citet{Chabrier2019}. These can be useful if the thermodynamic basis is not $(p,T)$.} two relatively simple ways modelling adiabatic interiors: using the tabulated values for $\nabla_{\rm{ad}}$, or calculating it using the triple-product rule 

\begin{equation}
    \left.\frac{\partial \log T}{\partial \log P} \right\vert_{S} = - \left.\frac{\partial \log S}{\partial \log P} \right\vert_{T} \left.\frac{\partial \log S}{\partial \log T}\right\vert_{P}^{-1} \, .
    \label{eq:triple_product}
\end{equation}

\noindent
Since the partial derivatives of the entropy are tabulated in CD21, they can be used directly, or be calculated from entropy derivatives from finite differencing or splines (when using the $(p, T)$ thermodynamic basis). Alternatively, assuming an ideal mixture, the adiabatic gradient of a hydrogen-helium mixture can be calculated from the pure hydrogen and helium tables and using the entropy of the mixture $S = X S_X + YS_Y + S_{\rm{mix}}$, where $X$, $Y$, $S_X$, and $S_Y$ are the mass fractions and entropy of hydrogen and helium, and $S_{\rm{mix}}$ is the mixing entropy.

\section{Results}\label{sec:results}

In this section we performed several numerical experiments with \textsc{Python} to determine the influence the previously mentioned numerical details have on the interiors of giant planets. We assumed a proto-solar hydrogen-helium mixture and evaluated the equations of state tables using cubic splines with \textsc{scipy.interpolate.RegularGridInterpolator}. The temperature equation was solved with \textsc{scipy.integrate.solve\_ivp} using the explicit fifth-order Runge-Kutta method \textsc{RK45} (additional solvers were tested in Appendix \ref{sec:solvers}), and the structure equations with a relaxation method \citep{Thorngren2016,mankovich_cassini_2019,2020ApJ...889...51M}. We assumed a planetary mass of 1 M$_{\rm{J}}$ and the surface boundary conditions were $T_{\rm{surf}}$ = 165 K at a pressure of $10^6$ Ba (1 bar). Additional calculations for different surface temperatures are in Appendix \ref{sec:surface_temperatures}.

\subsection{Differences due to the details of the adiabatic gradient}\label{sec:adiabatic_gradient}

We first investigated the influence of the form of the adiabatic gradient by calculating it in the following ways:

\begin{enumerate}
    \item Tabulated: Use the tabulated values of $\nabla_{\rm{ad}}$.
    \item Triple-product: Use Equation \ref{eq:triple_product} with the tabulated values of the entropy derivatives.
    \item Spline-derivatives: Use Equation \ref{eq:triple_product} with the entropy derivatives from the cubic spline.
    \item Finite-difference: Use Equation \ref{eq:triple_product} with the entropy derivatives from finite differencing.
    \item Ideal-mixing: Calculate the entropy of the mixture from the pure hydrogen and helium tables, use Equation \ref{eq:triple_product}, and evaluate the entropy derivatives from the cubic splines of the pure substances.
\end{enumerate}

\noindent
For the first four methods we used the pre-computed mixture tables of CD21. To establish a ground truth to serve as a baseline for the comparison, we used a root-finding algorithm to ensure a strictly isentropic interior profile  \texttt{scipy.optimize.elementwise.find\_root}.

\begin{figure}
    \centering
    \includegraphics[width=0.75\linewidth]{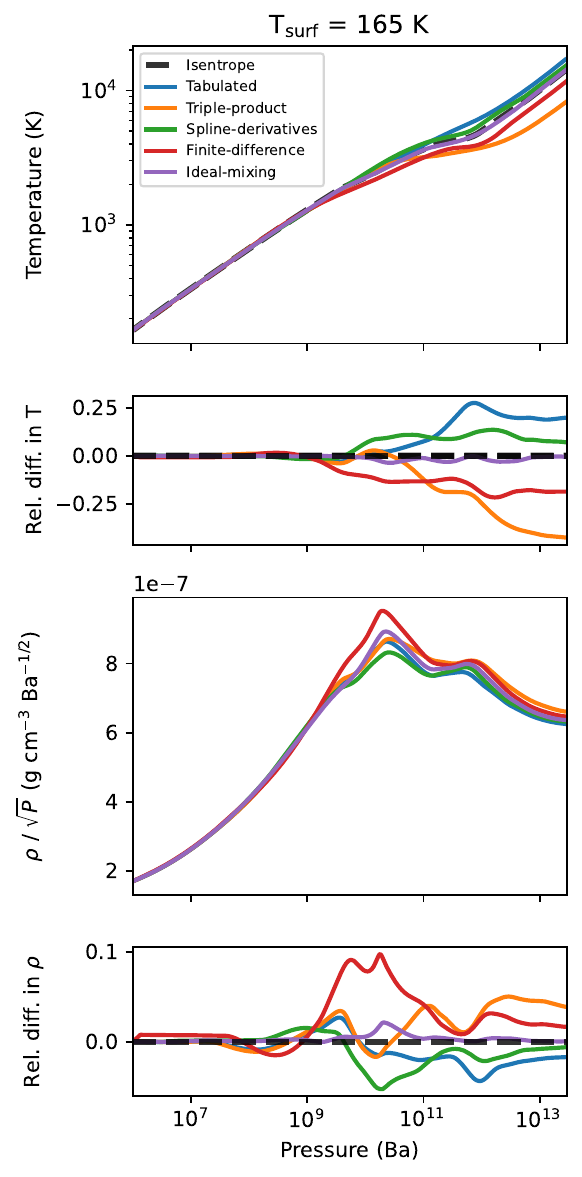}
    \caption{A comparison of the resulting adiabatic temperature and density profiles for different methods of calculating $\nabla_{\rm{ad}}$ (see legend and text for details). The dashed lines show the baseline profile calculated by enforcing an isentropic interior. The top two rows show the resulting temperature profiles and their relative difference compared to the baseline. The bottom two rows show $\rho/\sqrt{P}$ and the relative difference in the density compared to the baseline.}
    \label{fig:planet_solution_nabla_ad}
\end{figure}

The results are shown in Figure \ref{fig:planet_solution_nabla_ad}. The temperature and density profiles closely follow the isentropic profile until a pressure of about $10^9$ Ba ($10^3$ bar), where deviations start to increase and reach up to $\sim ^{+25\%}_{-40\%}$ in temperature or $^{+10\%}_{-5\%}$ in density. As a result, central temperatures can easily deviate by a few thousand K. Table \ref{tab:differences} lists the surface radii for the different methods and shows that they differ by up to 3.4\% compared to the isentropic solution. This is a significant difference: As a comparison, uncertainties of measured giant exoplanet sizes can be $\sim$1\% or even smaller.

Although it may be tempting to use these results to select the "best" method, in Appendix \ref{sec:surface_temperatures} we show that it depends on the surface temperature which method comes closest to the isentropic profile. These results demonstrate that the method of calculating the adiabatic gradient significantly influences the characterisation of giant planets interiors. We next examine how these results change when using the non-logarithmic form of the temperature equation.

\begin{table}
    \caption{Comparison of the surface radius for different methods to calculate the adiabatic gradient and using the logarithmic or non-logarithmic form of the temperature equation. Relative differences are with respect to the baseline derived from an isentropic interior. See text for details.}
    \begin{tabular}{c|cc}
        & \multicolumn{2}{c}{\textbf{Radius (R$_{\rm{J}}$) (rel. diff.)}} \\ \hline
        Isentrope & \multicolumn{2}{c}{1.022} \\ \hline
        \textbf{Method for $\nabla_{\rm{ad}}$} & Logarithmic & Non-logarithmic \\ \hline
        Tabulated & 1.041 (1.9\%) & 1.017 (-0.5\%) \\
        Triple-product & 0.987 (-3.4\%) & 1.014 (-0.8\%) \\
        Spline-derivatives & 1.032 (1.0\%) & 1.022 (0.0\%) \\
        Finite-difference & 1.002 (-2.0\%) & 1.022 (0.0\%) \\
        Ideal-mixing & 1.020 (-0.2\%) & 1.021 (-0.1\%) \\
    \end{tabular}
    \label{tab:differences}
\end{table}

\subsection{Differences due to the form of the temperature equation}\label{sec:temperature_equation}

The previous results used Equation \ref{eq:diff_eq_log} to derive adiabatic temperature profiles. Alternatively, it can be used in its non-logarithmic form:

\begin{equation}
    \frac{\partial T}{\partial p} = \frac{T}{p}\nabla_{\rm{ad}} \, .
    \label{eq:diff_eq}
\end{equation}

\begin{figure}
    \centering
    \includegraphics[width=0.75\linewidth]{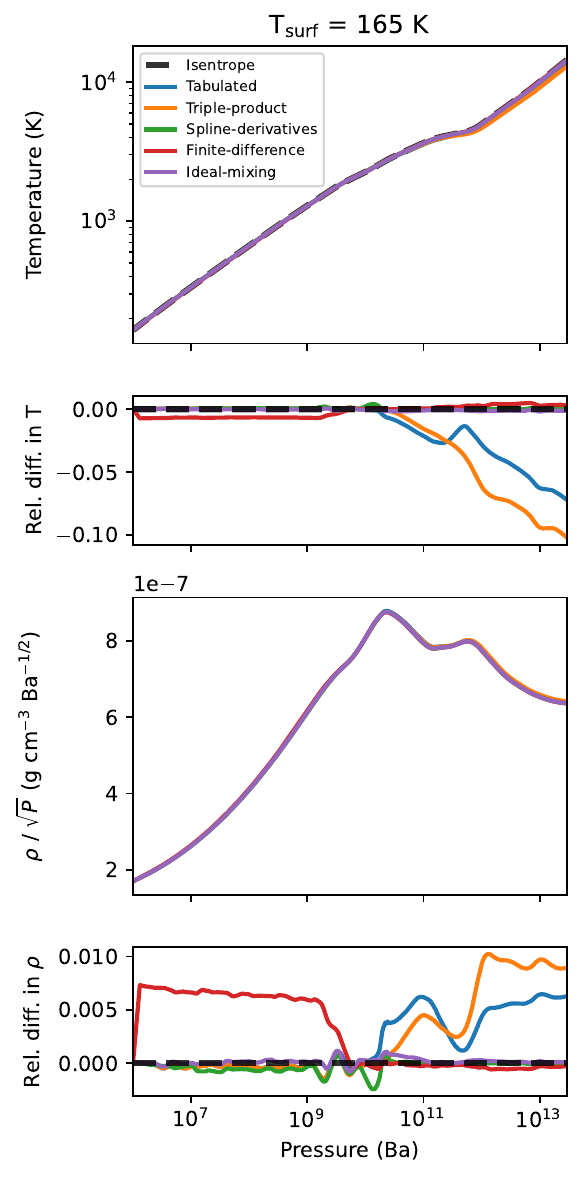}
    \caption{Same as Figure \ref{fig:planet_solution_nabla_ad}, but using the non-logarithmic version of the differential equation to determine the adiabatic temperature profile.}
    \label{fig:planet_solution_diff_eq}
\end{figure}

\noindent
We repeat the previous calculations, but solve Equation \ref{eq:diff_eq} (instead of Equation \ref{eq:diff_eq_log}). The inferred interior profiles are shown in Figure \ref{fig:planet_solution_diff_eq}, and the surface radii are listed in Table \ref{tab:differences}. Using the non-logarithmic version of the temperature equation reduces the deviations from the isentropic profile significantly. Now, the interior temperature and density deviate at most about -10\% or +1\% compared to the isentropic case. The different methods also produce a radius that is mostly consistent with the baseline, with differences up to $\sim$1\% for the worst-performing method.

\subsection{Practical recommendations}\label{sec:recommendations}

One way to avoid having to deal with these difficulties is to enforce an isentropic profile and use a root-finding procedure. However, while this is feasible for static models, it is too numerically expensive for evolution models since they require more evaluations and generally have a finer spatial resolution. If the adiabatic temperature gradient has to be used, our results suggest that the non-logarithmic temperature equation yields results closer matching an isentropic solution. In particular, additional calculations presented in Appendix \ref{sec:surface_temperatures} show that in this case there are methods that clearly perform much better or worse. Based on these results, we recommend using spline derivatives to calculate the adiabatic gradient with Equation \ref{eq:triple_product} and avoid using finite differences, the tabulated values of $\nabla_{\rm{ad}}$ or the entropy derivatives from the CD21 tables.

In this work we focused on the effect of the numerical details on the most widely used hydrogen-helium equation of state. We note that \citet{howard_jupiters_2023} previously compared interpolation schemes and found that they affect the models, and  also suggested that the pressure-temperature grid of the CD21 tables likely should be finer in particular regions of the parameter space. Additionally, our results suggest that these tables are not
thermodynamically consistent. The details of the comparisons likely would have changed when other tables \citep[e.g.,][]{Saumon1995,Becker2014}. However, the main goal of the models presented here is to highlight that there are significant effects depending on the employed numerical scheme. Furthermore, our main recommendation to enforce an isentropic profile whenever possible should apply to all tabulated equations of state.

\section{Summary and Conclusions}\label{sec:conclusions}

In this work, we investigated how the inferred internal structure and radius of convective giant planets depend on the numerical details of computing adiabatic temperature profiles. Using static interior models of a 1 M$_{\rm{J}}$ planet with widely used hydrogen-helium equation of state, we systematically compared different methods for evaluating the adiabatic temperature gradient and examined the effect of the form of the temperature structure equation. The main results from these numerical experiments are:

\begin{itemize}
    \item The interior and radius and of a planet are very sensitive to the numerical details of how adiabatic temperature profiles are calculated. 
    \item If numerically feasible, we recommend enforcing isentropic interiors instead of using the adiabatic temperature gradient. 
    \item Interior and evolution models should calculate $\nabla_{\rm{ad}}$ from spline derivatives and not use finite differences or tabulated values. We also recommend using the non-logarithmic version of the differential equation to derive the temperature profile when possible.
\end{itemize}

\noindent
These findings are particularly relevant in light of the ever-improving precision of giant exoplanet radius measurements, where observational uncertainties can reach $\sim$1\%. As interior and evolution models are  used to characterise an increasing number of exoplanets and probe their thermal histories, ensuring numerical consistency in the treatment of adiabatic profiles is essential for drawing robust physical conclusions.

\section*{Acknowledgements}
    We thank Christopher Mankovich for helpful discussions. We acknowledge support from the Swiss National Science Foundation (SNSF) grant \texttt{\detokenize{200020_215634}} and the National Centre for Competence in Research ‘PlanetS’ supported by SNSF. Extensive use was also made of the \textsc{Python} packages \textsc{Jupyter} \citep{jupyter}, \textsc{Matplotlib} \citep{Hunter2007}, \textsc{NumPy} \citep{harris2020array}, \textsc{SciPy} \citep{2020SciPy-NMeth}, and \textsc{tinyeos} \citep{tinyeos}.

\section*{Data Availability}
The code and all interiors models in this work will be made available
upon reasonable request to the authors.



\bibliographystyle{mnras}
\bibliography{library} % if your bibtex file is called example.bib

\begin{thebibliography}{}
\makeatletter
\relax
\def\mn@urlcharsother{\let\do\@makeother \do\$\do\&\do\#\do\^\do\_\do\%\do\~}
\def\mn@doi{\begingroup\mn@urlcharsother \@ifnextchar [ {\mn@doi@} {\mn@doi@[]}}
\def\mn@doi@[#1]#2{\def\@tempa{#1}\ifx\@tempa\@empty \href {http://dx.doi.org/#2} {doi:#2}\else \href {http://dx.doi.org/#2} {#1}\fi \endgroup}
\def\mn@eprint#1#2{\mn@eprint@#1:#2::\@nil}
\def\mn@eprint@arXiv#1{\href {http://arxiv.org/abs/#1} {{\tt arXiv:#1}}}
\def\mn@eprint@dblp#1{\href {http://dblp.uni-trier.de/rec/bibtex/#1.xml} {dblp:#1}}
\def\mn@eprint@#1:#2:#3:#4\@nil{\def\@tempa {#1}\def\@tempb {#2}\def\@tempc {#3}\ifx \@tempc \@empty \let \@tempc \@tempb \let \@tempb \@tempa \fi \ifx \@tempb \@empty \def\@tempb {arXiv}\fi \@ifundefined {mn@eprint@\@tempb}{\@tempb:\@tempc}{\expandafter \expandafter \csname mn@eprint@\@tempb\endcsname \expandafter{\@tempc}}}

\bibitem[\protect\citeauthoryear{{Becker}, {Lorenzen}, {Fortney}, {Nettelmann}, {Sch{\"o}ttler}  \& {Redmer}}{{Becker} et~al.}{2014}]{Becker2014}
{Becker} A.,  {Lorenzen} W.,  {Fortney} J.~J.,  {Nettelmann} N.,  {Sch{\"o}ttler} M.,   {Redmer} R.,  2014, \mn@doi [\apjs] {10.1088/0067-0049/215/2/21}, \href {https://ui.adsabs.harvard.edu/abs/2014ApJS..215...21B} {215, 21}

\bibitem[\protect\citeauthoryear{{B{\"o}hm-Vitense}}{{B{\"o}hm-Vitense}}{1958}]{1958ZA.....46..108B}
{B{\"o}hm-Vitense} E.,  1958, \zap, \href {https://ui.adsabs.harvard.edu/abs/1958ZA.....46..108B} {46, 108}

\bibitem[\protect\citeauthoryear{Burrows, Hubbard, Lunine  \& Liebert}{Burrows et~al.}{2001}]{burrows_theory_2001}
Burrows A.,  Hubbard W.~B.,  Lunine J.~I.,   Liebert J.,  2001, \mn@doi [Reviews of Modern Physics] {10.1103/RevModPhys.73.719}, 73, 719

\bibitem[\protect\citeauthoryear{{Chabrier} \& {Debras}}{{Chabrier} \& {Debras}}{2021}]{2021ApJ...917....4C}
{Chabrier} G.,  {Debras} F.,  2021, \mn@doi [\apj] {10.3847/1538-4357/abfc48}, \href {https://ui.adsabs.harvard.edu/abs/2021ApJ...917....4C} {917, 4}

\bibitem[\protect\citeauthoryear{{Chabrier}, {Mazevet}  \& {Soubiran}}{{Chabrier} et~al.}{2019}]{Chabrier2019}
{Chabrier} G.,  {Mazevet} S.,   {Soubiran} F.,  2019, \mn@doi [\apj] {10.3847/1538-4357/aaf99f}, \href {https://ui.adsabs.harvard.edu/abs/2019ApJ...872...51C} {872, 51}

\bibitem[\protect\citeauthoryear{{Fortney} \& {Nettelmann}}{{Fortney} \& {Nettelmann}}{2010}]{Fortney2010}
{Fortney} J.~J.,  {Nettelmann} N.,  2010, \mn@doi [\ssr] {10.1007/s11214-009-9582-x}, \href {https://ui.adsabs.harvard.edu/abs/2010SSRv..152..423F} {152, 423}

\bibitem[\protect\citeauthoryear{{Guillot}}{{Guillot}}{1999}]{Guillot1999}
{Guillot} T.,  1999, \mn@doi [Science] {10.1126/science.286.5437.72}, \href {https://ui.adsabs.harvard.edu/abs/1999Sci...286...72G} {286, 72}

\bibitem[\protect\citeauthoryear{{Guillot}}{{Guillot}}{2005}]{Guillot2005}
{Guillot} T.,  2005, \mn@doi [Annual Review of Earth and Planetary Sciences] {10.1146/annurev.earth.32.101802.120325}, \href {https://ui.adsabs.harvard.edu/abs/2005AREPS..33..493G} {33, 493}

\bibitem[\protect\citeauthoryear{{Hansen} \& {Kawaler}}{{Hansen} \& {Kawaler}}{1994}]{1994sipp.book.....H}
{Hansen} C.~J.,  {Kawaler} S.~D.,  1994, {Stellar Interiors. Physical Principles, Structure, and Evolution.}, \mn@doi{10.1007/978-1-4419-9110-2.
}

\bibitem[\protect\citeauthoryear{Harris et~al.,}{Harris et~al.}{2020}]{harris2020array}
Harris C.~R.,  et~al., 2020, \mn@doi [Nature] {10.1038/s41586-020-2649-2}, 585, 357

\bibitem[\protect\citeauthoryear{Helled \& Howard}{Helled \& Howard}{2026}]{helled_giant_2026}
Helled R.,  Howard S.,  2026, in Encyclopedia of Astrophysics. eprint: arXiv:2407.05853, pp 51--65, \mn@doi{10.1016/B978-0-443-21439-4.00013-4}, \url {https://ui.adsabs.harvard.edu/abs/2026enap....1...51H}

\bibitem[\protect\citeauthoryear{{Helled} et~al.,}{{Helled} et~al.}{2014}]{Helled2014}
{Helled} R.,  et~al., 2014, in {Beuther} H.,  {Klessen} R.~S.,  {Dullemond} C.~P.,   {Henning} T.,  eds, Protostars and Planets VI. University of Arizona Press, p.~643 (\mn@eprint {arXiv} {1311.1142}), \mn@doi{10.2458/azu_uapress_9780816531240-ch028}

\bibitem[\protect\citeauthoryear{{Henyey}, {Vardya}  \& {Bodenheimer}}{{Henyey} et~al.}{1965}]{Henyey1965}
{Henyey} L.,  {Vardya} M.~S.,   {Bodenheimer} P.,  1965, \mn@doi [\apj] {10.1086/148357}, \href {https://ui.adsabs.harvard.edu/abs/1965ApJ...142..841H} {142, 841}

\bibitem[\protect\citeauthoryear{Howard et~al.,}{Howard et~al.}{2023}]{howard_jupiters_2023}
Howard S.,  et~al., 2023, \mn@doi [Astronomy and Astrophysics] {10.1051/0004-6361/202245625}, 672, A33

\bibitem[\protect\citeauthoryear{Hunter}{Hunter}{2007}]{Hunter2007}
Hunter J.~D.,  2007, \mn@doi [Computing in Science \& Engineering] {10.1109/MCSE.2007.55}, 9, 90

\bibitem[\protect\citeauthoryear{{Kippenhahn}, {Weigert}  \& {Weiss}}{{Kippenhahn} et~al.}{2012}]{Kippenhahn2012}
{Kippenhahn} R.,  {Weigert} A.,   {Weiss} A.,  2012, {Stellar Structure and Evolution}.
Springer-Verlag Berlin Heidelberg, \mn@doi{10.1007/978-3-642-30304-3}

\bibitem[\protect\citeauthoryear{Kluyver et~al.,}{Kluyver et~al.}{2016}]{jupyter}
Kluyver T.,  et~al., 2016, in Loizides F.,  Scmidt B.,  eds, Positioning and Power in Academic Publishing: Players, Agents and Agendas. IOS Press, Netherlands, pp 87--90, \url {https://eprints.soton.ac.uk/403913/}

\bibitem[\protect\citeauthoryear{Lamers \& Levesque}{Lamers \& Levesque}{2017}]{lamers_understanding_2017}
Lamers H. J. G. L.~M.,  Levesque E.~M.,  2017, Understanding {Stellar} {Evolution}.
\url {https://ui.adsabs.harvard.edu/abs/2017use..book.....L}

\bibitem[\protect\citeauthoryear{{Mankovich} \& {Fortney}}{{Mankovich} \& {Fortney}}{2020}]{2020ApJ...889...51M}
{Mankovich} C.~R.,  {Fortney} J.~J.,  2020, \mn@doi [\apj] {10.3847/1538-4357/ab6210}, \href {https://ui.adsabs.harvard.edu/abs/2020ApJ...889...51M} {889, 51}

\bibitem[\protect\citeauthoryear{Mankovich, Marley, Fortney  \& Movshovitz}{Mankovich et~al.}{2019}]{mankovich_cassini_2019}
Mankovich C.,  Marley M.~S.,  Fortney J.~J.,   Movshovitz N.,  2019, \mn@doi [The Astrophysical Journal] {10.3847/1538-4357/aaf798}, 871, 1

\bibitem[\protect\citeauthoryear{{M{\"u}ller} \& {Helled}}{{M{\"u}ller} \& {Helled}}{2023}]{Muller2023}
{M{\"u}ller} S.,  {Helled} R.,  2023, \mn@doi [Frontiers in Astronomy and Space Sciences] {10.3389/fspas.2023.1179000}, \href {https://ui.adsabs.harvard.edu/abs/2023FrASS..1079000M} {10, 1179000}

\bibitem[\protect\citeauthoryear{Müller}{Müller}{2025}]{tinyeos}
Müller S.,  2025, tinyeos, \url{https://github.com/tiny-hippo/tinyeos}

\bibitem[\protect\citeauthoryear{{Saumon}, {Chabrier}  \& {van Horn}}{{Saumon} et~al.}{1995}]{Saumon1995}
{Saumon} D.,  {Chabrier} G.,   {van Horn} H.~M.,  1995, \mn@doi [\apjs] {10.1086/192204}, \href {https://ui.adsabs.harvard.edu/abs/1995ApJS...99..713S} {99, 713}

\bibitem[\protect\citeauthoryear{Schwarzschild}{Schwarzschild}{1906}]{schwarzschild_equilibrium_1906}
Schwarzschild K.,  1906, Nachrichten von der Königlichen Gesellschaft der Wissenschaften zu Göttingen. Math.-phys. Klasse, 195, 41

\bibitem[\protect\citeauthoryear{{Stevenson}}{{Stevenson}}{1982}]{Stevenson1982}
{Stevenson} D.~J.,  1982, \mn@doi [Annual Review of Earth and Planetary Sciences] {10.1146/annurev.ea.10.050182.001353}, \href {https://ui.adsabs.harvard.edu/abs/1982AREPS..10..257S} {10, 257}

\bibitem[\protect\citeauthoryear{{Thorngren} \& {Fortney}}{{Thorngren} \& {Fortney}}{2019}]{2019ApJ...874L..31T}
{Thorngren} D.,  {Fortney} J.~J.,  2019, \mn@doi [\apjl] {10.3847/2041-8213/ab1137}, \href {https://ui.adsabs.harvard.edu/abs/2019ApJ...874L..31T} {874, L31}

\bibitem[\protect\citeauthoryear{{Thorngren}, {Fortney}, {Murray-Clay}  \& {Lopez}}{{Thorngren} et~al.}{2016}]{Thorngren2016}
{Thorngren} D.~P.,  {Fortney} J.~J.,  {Murray-Clay} R.~A.,   {Lopez} E.~D.,  2016, \mn@doi [\apj] {10.3847/0004-637X/831/1/64}, \href {https://ui.adsabs.harvard.edu/abs/2016ApJ...831...64T} {831, 64}

\bibitem[\protect\citeauthoryear{Virtanen et~al.,}{Virtanen et~al.}{2020}]{2020SciPy-NMeth}
Virtanen P.,  et~al., 2020, \mn@doi [Nature Methods] {10.1038/s41592-019-0686-2}, \href {https://rdcu.be/b08Wh} {17, 261}

\makeatother
\end{thebibliography}

% \clearpage
\appendix
% \nolinenumbers  % remove line numbers for arXiv

\section{The influence of the numerical solver on adiabatic interiors}\label{sec:solvers}

In this section, we investigate how the choice of numerical solver for Equation \ref{eq:diff_eq_log} affects the adiabatic interiors in static interior models. As in Section \ref{sec:results}, the differential equation was solved with \textsc{scipy.integrate.solve\_ivp}, but we varied the integration methods. The adiabatic gradient was calculated from the \citet{2021ApJ...917....4C} tables assuming an ideal mixture $S = X S_X + YS_Y + S_{\rm{mix}}$ and using Equation \ref{eq:triple_product}, where the derivatives were calculated from cubic splines.

\begin{figure}
    \centering
    \includegraphics[width=0.75\linewidth]{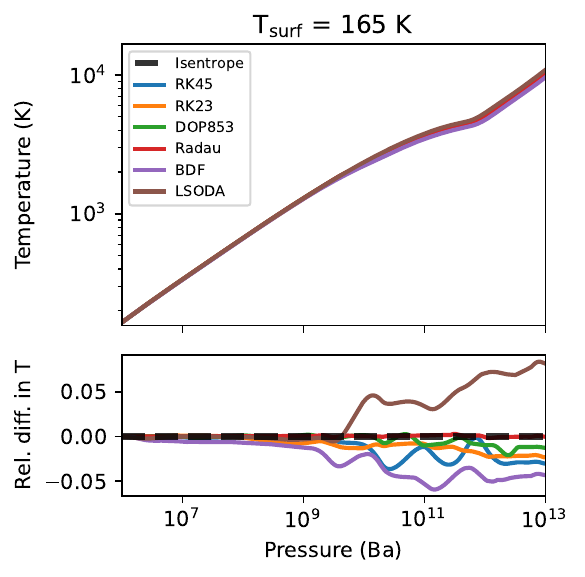}
    \caption{Comparison of interior profiles for different integration methods of Equation \ref{eq:diff_eq_log}. The baseline isentropic profiles are shown in dashed black lines. See text for details.}
    \label{fig:solver_comparison}
\end{figure}

The results are shown in Figure \ref{fig:solver_comparison}. The resulting interiors are also influenced by the integration method, but less so compared to the method of calculating the adiabatic gradient or the form of the differential equation. While for this specific case the explicit solvers (RK45, RK23, DOP853) performed better than implicit solvers (Radau, BDF, LSODA), additional calculations showed that this is not always the case and depends on the planetary parameters. 

\section{Adiabatic models for different surface temperatures}\label{sec:surface_temperatures}

In this section, we repeat the calculations from Sections \ref{sec:adiabatic_gradient} and \ref{sec:temperature_equation} for different surface temperatures, namely $T_{\rm{surf}} = 250, 500, 750$ and $1000 K$. This covers a large range potential surface temperatures of young and old planets as well as weakly to strongly irradiated ones.

\begin{figure}
    \begin{subfigure}[b]{\linewidth}
        \centering
        \includegraphics[width=0.6\linewidth]{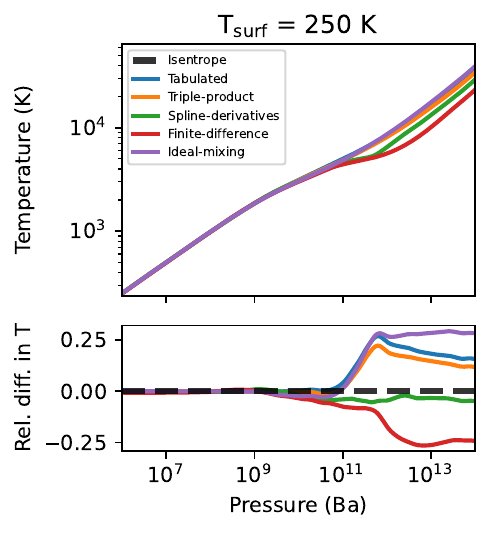}
    \end{subfigure} 
    \begin{subfigure}[b]{\linewidth}
        \centering
        \includegraphics[width=0.6\linewidth]{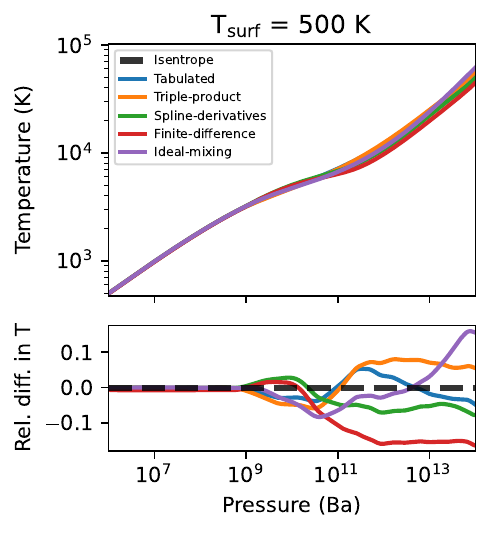}
    \end{subfigure}
    \begin{subfigure}[b]{\linewidth}
        \centering
        \includegraphics[width=0.6\linewidth]{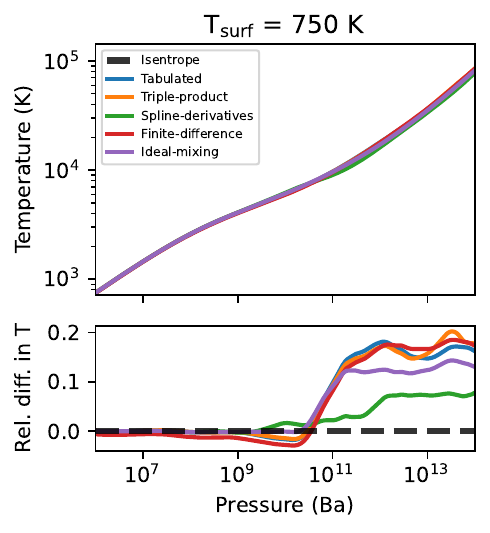}
    \end{subfigure}
    \begin{subfigure}[b]{\linewidth}
        \centering
        \includegraphics[width=0.6\linewidth]{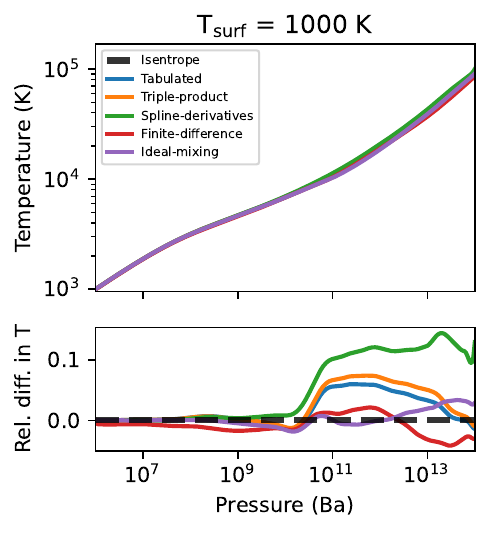}
    \end{subfigure}
    \caption{Comparison of interior profiles when solving the logarithmic temperature equation (Eq. \ref{eq:diff_eq_log}) for different method of calculating the adiabatic gradient (see legend and Section \ref{sec:results}). The baseline isentropic profiles are shown in dashed black lines. There is no clear best-performing method since it changes depending on the surface temperature.}
    \label{fig:nabla_ad_comparison}
\end{figure}

\begin{figure}
    \begin{subfigure}[b]{\linewidth}
        \centering
        \includegraphics[width=0.6\linewidth]{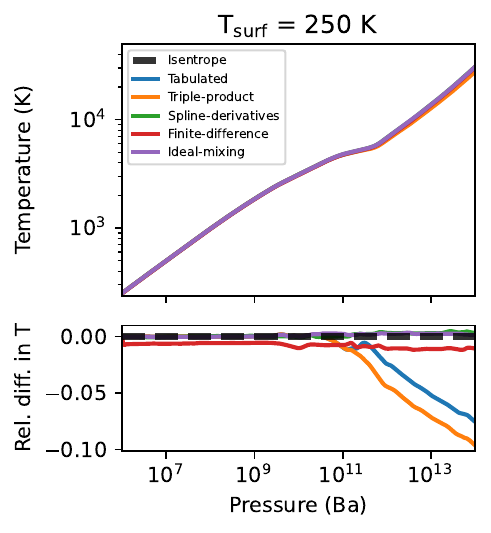}
    \end{subfigure}
    \begin{subfigure}[b]{\linewidth}
        \centering
        \includegraphics[width=0.6\linewidth]{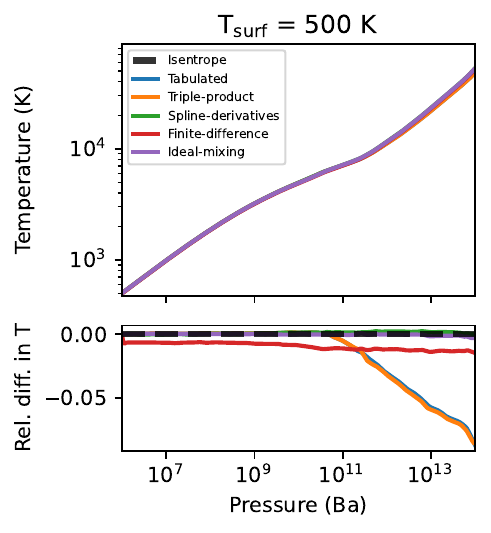}
    \end{subfigure}
    \begin{subfigure}[b]{\linewidth}
        \centering
        \includegraphics[width=0.6\linewidth]{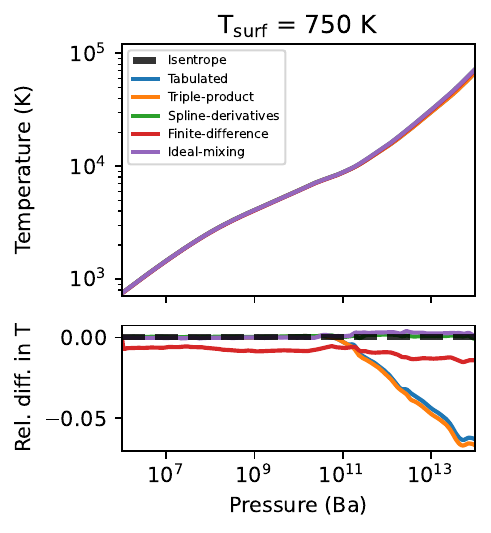}
    \end{subfigure}
    \begin{subfigure}[b]{\linewidth}
        \centering
        \includegraphics[width=0.6\linewidth]{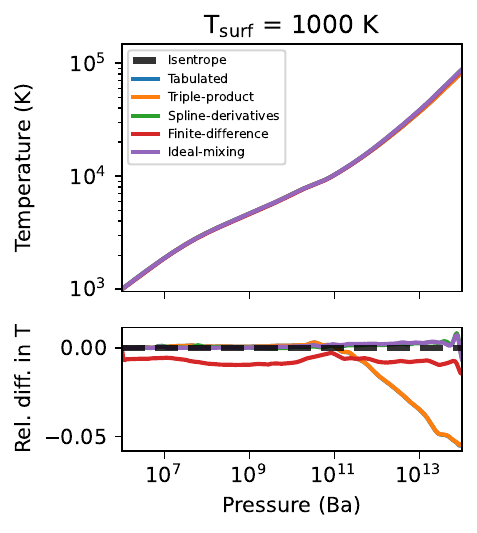}
    \end{subfigure}
    \caption{Comparison of interior profiles when solving the non-logarithmic temperature equation (Eq. \ref{eq:diff_eq}) for different method of calculating the adiabatic gradient (see legend and Section \ref{sec:results}). In contrast to the results in Figure \ref{fig:nabla_ad_comparison}, there are now methods that clearly perform worse independent of the surface temperature.}
    \label{fig:nabla_ad_comparison_nonlog}
\end{figure}

The resulting interior profiles when we solved the logarithmic temperature equation are shown in Figure \ref{fig:nabla_ad_comparison}. Surprisingly, no clear winner emerges as to which method best matches the ground truth from the isentropic profile. It is rather the case that the best method depends on the surface temperature. We repeated these calculations but used the non-logarithmic temperature equation, and the results are shown in Figure \ref{fig:nabla_ad_comparison_nonlog}. This time, the trends are stable across the tested surface temperatures: The worst-performing methods are clearly using finite differences, or the tabulated values of the adiabatic gradient and the entropy derivatives. Based on these results, we therefore generally recommend to use spline derivatives.

% Don't change these lines
\bsp	% typesetting comment
\label{lastpage}
\end{document}